\documentstyle[preprint,aps,prbbib,psfig]{revtex}

\begin{document}
\draft
\preprint{ }

\title{Universal Phase Diagram for Vortex States of Layered  
Superconductors in Strong Magnetic Fields}

\author{Jun Hu$^{1,2}$ and A.H. MacDonald$^{2}$}

\address{$^{1}$ National High Magnetic Field Laboratory,
Florida State University, Tallahassee FL 32306} 

\address{$^2$ Department of Physics, Indiana University,
 Bloomington IN 47405}

\date{\today}
\maketitle

\begin{abstract}

We report on a Monte Carlo study of the lowest-Landau-level limit 
of the Lawrence-Doniach model for a layered superconductor.
We have studied order parameter correlation functions 
for indications of the broken translational symmetry and the off-diagonal
long range order present in the mean-field-theory vortex
lattice.  Our results are consistent with a single first order 
phase transition between a low temperature 3D vortex solid phase,
with both broken translational symmetry and off-diagonal long range order,
and a high temperature vortex liquid phase with 
no broken symmetries.  We construct a universal 
phase diagram in terms of dimensionless parameters characterizing
intra-layer and inter-layer couplings. 
The universal phase boundary extracted from our simulations
and the associated latent heats and magnetization jumps  
are compared with experiment and with numerical results obtained for 
related models.

\end{abstract}


\section{Introduction} 

Because of the combination of high transition temperature,
strong planar anisotropy, and short coherence length that occurs in high
temperature superconductors, strong thermal fluctuations are present
over a wide temperature interval in these materials.
Thermal fluctuations are especially important in a magnetic field
where they are responsible for the melting 
of the Abrikosov\cite{abrikosov} vortex lattice, often at temperatures
that are well below the mean-field critical temperature, giving rise to
a vortex liquid 
state~\cite{liquid,fisher,junhuliquid,nikulov,exp1,exp2,exp3,exp4,exp5} 
that is strongly diamagnetic
but also strongly resistive.  In this paper\cite{junhuthesis} we report 
on a Monte Carlo study of thermal fluctuations in layered superconductors 
with a magnetic field oriented perpendicular to the planes.  Our study 
is based on a Lawrence-Doniach model\cite{ldmodel} and employs the
lowest-Landau-level (LLL) approximation\cite{lll,sorensen,friesen}
which is valid near the mean-field-transition temperature.
The emphasis is on generic properties of this model rather than on quantitative 
estimates of phase diagrams, latent heats, and magnetization discontinuities
for particular materials.\cite{stroud}
We find that, for this model, a single first order phase transition
occurs between a low-temperature state and a high temperature 
state with no broken symmetries.  The low temperature state 
has off-diagonal long-range-order (ODLRO) 
both along and perpendicular to the field direction
and broken translational symmetry.

At the mean-field level, the phase
transition between the normal state and the Abrikosov lattice state is 
highly unusual in two related respects.   Firstly, the 
eigenvalue of the Cooper-pair 
density matrix which diverges at the transition point 
has a macroscopic degeneracy, 
in contrast to the isolated divergent eigenvalue 
with a zero-momentum-state eigenfunction found at zero magnetic field.
The many divergent eigenvalues are 
associated with the many states in the Landau levels\cite{cpll} 
into which the transverse translational degrees of freedom of the 
Cooper pairs are quantized.  At the mean-field level, this unusual
property is responsible for the simultaneous development of superconducting
order and broken translational symmetry in the Abrikosov vortex 
lattice state.  Another consequence is that, at temperatures 
well above the mean-field transition temperature
$T_c^{MF}$, fluctuations in a magnetic field in $D$-dimensions
are like those of a $D-2$ dimensional
system\cite{leeshenoy} at zero magnetic field.
Br\'{e}zin {\it et al.}\cite{brezin} first suggested, 
on the basis of expansions  around the upper critical dimension
for this problem ($D=6$), that fluctuations would drive 
the Abrikosov transition first order.
Indeed Hetzel {\it et al.}\cite{hetzel} some time ago found evidence 
from Monte Carlo simulations that the transition is first order for $D=3$.
This conclusion has been substantiated by a large
volume of subsequent\cite{subsequent} work from various points
of view, although there is not yet universal agreement 
on the nature of the fluctuation altered phase diagram.  
In particular, some workers doubt the occurrence of any true phase transition
in a magnetic field.~\cite{moore}  
Others have suggested that the temperature $T_c$
at which superconductivity along the field first occurs  
is either larger~\cite{chenteitel,sudbo} or smaller~\cite{sudbo,glazman}
than the temperature $T_M$ where translational symmetry is broken  
and the vortex lattice forms.  Here we find that, when
the LLL approximation is valid, $T_c = T_M$ and the transition 
is first order.

The LLL approximation is valid near the mean-field transition 
temperature.  However, 
fluctuations in many high temperature superconductors 
are strong enough to drive the 
transition field outside of its range of validity,\cite{validitycaveat} 
at least at the high temperatures where the transition field strengths 
can be reached in standard superconducting magnets.
The LLL approximation is, by conservative estimates, accurate 
for fields exceeding half of the mean field $H_{c2}$.  When fluctuations in 
higher Landau levels are still weak, their presence can be accounted for 
by renormalizing the parameters of the model.~\cite{lll,thouless}
Partly for this reason, it has proved difficult to
determine the range of validity of the LLL
approximation by comparing with experimental data; for example 
the minimum field above which the LLL approximation is accurate in YBCO 
was estimated to be $\sim 10$ {\rm Tesla} in one recent study\cite{friesen}
and $\sim 2 $ {\rm Tesla} in another.\cite{pierson}
A characteristic property of the LLL limit of the Lawrence-Doniach
model is that the number of vortices passing through 
each plane is fixed by the magnetic field strength and does 
not fluctuate.  The model must therefore fail qualitatively
at very weak fields since, in the limit of zero magnetic field,
thermally generated vortex loops are 
believed to play an essential role near the superfluid transition.
We will return to this problematic issue in discussing the results
of our simulations.  
 
We have previously reported on a simulation of LLL 
thermodynamics for two-dimensional electron systems.\cite{2d}  
The present paper reports on simulations of the 
Josephson coupled layers that constitute the 
Lawrence-Doniach model for superconductors with strong planar anisotropy.
In Section II we summarize and discuss the LLL limit of the 
Lawrence-Doniach model and introduce the correlation functions
whose temperature dependence we have studied.  In Section
III we present our simulation results.  In Section 
IV we discuss some implications of our  
simulations, commenting on their relationship to experiment and to 
other simulations.  Finally we briefly summarize our study 
in Section V. 

\section{LLL Limit of the Lawrence-Doniach Model}  

In the Lawrence-Doniach model,\cite{ldmodel} the local superconducting
order parameter $\Psi_n(\vec r)$ is defined on discretely labeled 
continuum layers.
(Here and in what follows $\vec r$ is a two-dimensional (2D) coordinate, 
taken to be the $\hat x- \hat y$ plane, and 
$n$ is a layer index.) The Ginzburg-Landau free energy for a 
particular configuration of the order parameter is given by 
\begin{equation}
F_{GL} = d_0 \sum_n \int d^2 \vec r f[\Psi_n]
\label{eq:one}
\end{equation}
where 
\begin{eqnarray}
f[\Psi_n]  & = & \sum_n [\alpha (T) \vert \Psi_n(\vec r)
 \vert^2 + \frac{\beta}{2} \vert \Psi_n(\vec r) \vert^4 
    + \frac{1}{2m_{ab}}\vert
 (-i \hbar \nabla - 2e \vec A)\Psi_n(\vec r) \vert^2 + \nonumber \\
 & & + \frac{\hbar^2}{2m_cd^2} \vert \Psi_{n+1}(\vec r)\exp
 (\frac{2\pi i}{\Phi_0}\int_{nd}^{(n+1)d}A_zdz)-\Psi_n(\vec r) \vert^2],
\label{eq:two}
\end{eqnarray}
$d_0$ is the layer thickness, and 
$d$ is the distance between layers.  In the limit 
where typical order parameter functions vary smoothly as a function of 
layer index this reduces to a three-dimensional ($3D$) Ginzburg-Landau
model with anisotropy\cite{anisoglreview} parameter $\gamma^2 = m_c / m_{ab}$. 
The mean field transition temperature 
in the absence of a field ($T_{c0}^{MF}$)
occurs where $\alpha (T)$ vanishes.  Where definiteness is
required we will follow the usual 
practice of assuming that $\alpha$ varies linearly with temperature
and that other parameters of the model are temperature independent.
(This assumption is unlikely to be valid across the entire 
fluctuation regime in typical high-temperature superconductors.)  
We limit our attention to magnetic fields directed perpendicular 
to the layers and choose a  Landau gauge with vector potential
$\vec A = (0, Hx, 0)$ where $H$ is the magnetic field
strength.\cite{fieldfluc}  The 2D gradient term in this energy functional
is diagonalized by kinetic energy eigenfunctions for particles of 
charge $2e$ in a perpendicular magnetic field.  In the 
underlying fermionic description these particles are the 
electronic Cooper pairs.  Sufficiently close to the mean-field
critical temperature, fluctuations are dominated by those in the 
Hilbert subspace of macroscopic dimension within which 
the Cooper pair kinetic energy is minimized.
Within this manifold, the 2D kinetic energy operator
can be replaced by a constant, $ \hbar e H / m_{ab} c$.  This replacement,
combined with the corresponding constraint on the order parameter,
is the LLL approximation.  

Our simulations are performed on finite systems consisting of 
$N_z$ rectangular layers with sides of length $L_x$ and $L_y$.
The quasiperiodic boundary conditions consistent with our gauge choice are: 
\begin{eqnarray}
\Psi_n (x,y + L_y) & = &  \Psi_n (x,y) \nonumber \\
\Psi_n (x + L_x, y) & = & \exp ( i 2 \pi N_{\phi} (\frac{y}{L_y})) 
                          \Psi_n (x,y) \nonumber \\
\Psi_{n+N_z} (x,y) & = & \Psi_n (x,y) 
\label{eq:three}
\end{eqnarray} 
Here $N_{\phi} = (L_x L_y) / (2 \pi \ell^2)$ is the number of flux quanta of
the internal field passing through each layer 
which must be an integer if the finite-size boundary conditions are to 
be satisfied. (The magnetic length $\ell^2 \equiv \hbar c/ 2 e H$.) 
In the LLL approximation,
the order parameter in each layer is expanded in terms of a 
complete set of minimal-gradient-energy eigenfunctions that satisfy
these boundary conditions.  It turns out that this 2D basis set
consists of $N_{\phi}$ displaced
elliptic theta functions.\cite{2d,haldanetrans}.  Therefore the order 
parameter is expanded in the form  
\begin{equation}
\Psi_n (\vec r) =  \big(\frac{|\alpha_H| \pi \ell^2}{\beta}\big)^{1/2}
\sum_j C_{n,j} [ \sum_{s = -\infty}^{\infty}  (L_y)^{-1/2}(\pi
\ell^2)^{-1/4} \exp (i y X_{j,s} /\ell^2 )
\exp (-(x-X_{j,s})^2/ 2 \ell^2)]
\label{eq:four}
\end{equation}
In Eq.~(\ref{eq:four}) $\alpha_H = \alpha (T) + (\hbar e H / m_{ab} c) 
= \alpha(T)
( 1 - H / H_{c2}^{MF}(T)) $,
$X_{j,s} = j 2 \pi \ell^2/ L_y + s L_x$,
$j$ runs from $1$ to $N_{\phi}$, and $n$ runs from $1$ to $N_z$.
The normalization is chosen so that the average value of $|c_j|^2$ 
is $\sim 1$ for Abrikosov's mean-field order parameter.
As we see explicitly in Eq.~(\ref{eq:four}), the order parameter in the 
LLL approximation in this gauge is the product of $\exp ( -x^2 /2 \ell^2 )$ 
and an analytic function of $z \equiv x + i y$.
This constraint, whose analog in another gauge has been emphasized in
the work of Te\v{s}anovi\'{c} and collaborators\cite{zlatkoearly} 
and powerfully employed in studies of the 
quantum Hall effect, means that amplitude and phase fluctuations of the 
order parameter are necessarily linked in this regime.  Since
the line integral of the phase gradient of an analytic complex-valued 
function around any closed loop
is $2 \pi$ times the number of enclosed zeroes, it follows from    
Eq.~(\ref{eq:three}) 
that for each layer index $n$ and each configuration of the
order parameter, $\Psi_n (\vec r)$ has precisely $N_{\phi}$ zeroes.
There are no thermally generated flux loops in the LLL approximation.
Moreover, for each $n$, $\Psi_n (\vec r)$ is specified up to a
complex overall factor by the positions of its zeroes.  

We will characterize the states that appear in our simulations in terms of 
intra-layer and inter-layer correlation functions constructed in terms 
of the quantity $\Delta_{n',n}(\vec q)$ defined by the following 
equation:
\begin{equation} 
\int d^2 \vec r \,\bar \Psi_{n^\prime} (\vec r)\Psi_n (\vec r) 
          \exp \{ - i \vec q \cdot \vec r \} = 
         \frac{ |\alpha_H| \pi \ell^2 }{\beta} 
N_\phi \Delta_{n^\prime,n}(\vec q) \exp \{-\frac{q^2\ell^2}{4} \}. 
\label{eq:five}
\end{equation}
In terms of order parameter expansion coefficients for the finite
system $\Delta_{n^\prime,n} (\vec q)$ has the form:
\begin{equation}
\Delta_{n^\prime,n} (\vec q)  = 
\frac{1}{N_\phi}\sum_{j_1j_2}\bar C_{n^\prime,j_1} C_{n,j_2}
\delta_{j_2-j_1-n_y}
\exp [-iq_x (X_{j_1}+X_{j_2})/2]
\label{eq:six} 
\end{equation}
where $\delta_j = 1$ if $j$ is a multiple of $N_\phi$ and
is zero otherwise, and $X_j \equiv X_{j,0}$.
For the Fourier expansions of our finite systems with 
periodic boundary conditions, $\vec q = 2\pi(n_x/L_x,n_y/L_y)$ 
where $n_x$ and $n_y$ vary from over any range of $N_{\phi}$ 
consecutive values. 
This quantity is conveniently sampled in our Landau gauge Monte Carlo 
simulations. 
$\Delta_{n^\prime,n}(\vec q)$ is proportional to the Fourier transform 
of an order parameter product that is diagonal in planar coordinates
but off-diagonal in layer indices.  We remark that because of the 
analyticity property of LLL wavefunctions this quantity nevertheless  
completely specifies the off-diagonal order parameter product.  In particular,
inverting the Fourier transform and using analyticity it can be  
shown that\cite{arovas,analyticity} 
\begin{eqnarray}
 \bar \Psi_{n^\prime} (\vec r')\Psi_n (\vec r) & = &
 \big(\frac{|\alpha_H|\pi\ell^2}{2\beta}\big)
 \exp ( - \frac{| \vec r - \vec r'|^2}{ 4 \ell^2}) 
 \exp ( \frac{i (xy - x'y')}{2 \ell^2})
 \sum_{\vec q} \Delta_{n',n}( \vec q) 
               \exp ( -\frac{q^2 \ell^2}{4}) \nonumber \\
& & \times \exp (\frac{i \vec q \cdot (\vec r + \vec r')}{2}) 
	\exp (\frac{\vec q \times (\vec r' - \vec r)}{2}).
\label{eq:seven}
\end{eqnarray} 

We will principally be interested in whether or not our simulations 
show evidence 
for broken translational symmetry in the spatial distribution of the local 
superfluid density $ |\Psi_n(\vec r)|^2$ or evidence for ODLRO
in the order parameter.  We define the quantity 
\begin{equation}
\Delta_0 = \frac{1}{N_z} \sum_n \Delta_{n,n} (\vec q = 0).  
\label{eq:eight}
\end{equation}
$\Delta_0$ is proportional to the average local 
superfluid density over all layers.
In mean-field theory $\Delta_0$ may be taken as the order parameter of the 
Abrikosov state.  (With our normalization, 
$\Delta_0 = 2 / \beta_{\bigtriangleup} $ where\cite{abref} 
$\beta_\bigtriangleup \approx 1.159595$ in
the Abrikosov state.)  When fluctuations are included, we expect that 
$\Delta_0$ will vary from zero well above the mean-field
transition temperature to $2 / \beta_\bigtriangleup $ at zero temperature,
with singular behavior at an intermediate temperature if a phase
transition occurs.
A non-zero value for $\Delta_0$ does not indicate either
broken translational symmetry 
or ODLRO.  Broken translational symmetry can be
identified by examining 
\begin{equation}
S_{n',n} (\vec q) \equiv  \frac{\Delta_{n^\prime,n^\prime}(-\vec q) 
							\Delta_{n,n}(\vec q)}{ \Delta_0^2}.
\label{eq:nine}
\end{equation} 
$S_{n',n}$ measures intra-layer ($n^\prime = n$ ) and 
inter-layer ($n^\prime \neq n$ ) 
superfluid density correlations.   In the Abrikosov mean-field solution 
$S_{n',n} (\vec q) = 1 $  when $\vec q$ equals a reciprocal lattice vector
($\vec G$) of the vortex lattice and is otherwise zero.  For states with broken
translation symmetry, the thermal average of $S_{n',n}(\vec G)$ will remain 
finite for $N_{\phi} \to \infty$.  For a vortex liquid state we expect that 
$S_{n',n}(\vec q)$ will vanish as $N_{\phi}^{-1}$ at large $N_{\phi}$
for all
wavevectors $\vec q$.  $S_{n',n}(\vec q)$ is tantamount to the static 
structure factor of the vortices, although this equivalence is not in general 
precise as far as we are aware.

Off-diagonal long-range order along the field 
can be identified by examining 
\begin{equation}
z \equiv \frac{1}{N_z^2}\sum_{n,n'}\frac{\Delta_{n',n}(\vec q = 0)}{\Delta_0}. 
\label{eq:ten}
\end{equation}
For the Abrikosov mean-field solution $z = 1$.  
For states with ODLRO along the field
the thermal average of $z$ will remain finite in the
thermodynamic limit, otherwise it should go to 
zero as $N_z^{-1}$.  Application of the notion of ODLRO
to directions perpendicular to the field
is more subtle and will be discussed later.

The Ginzburg-Landau free energy can be expressed in terms of
$ \Delta_{n^\prime,n} (\vec q) $: 
\begin{eqnarray}
\frac{F_{GL}}{k_BT} & = & N_{\phi} g^2 
    \sum_n \left[\rm{sgn} (\alpha_H) \Delta_{n,n}(0)
+ \frac{1}{4}\sum_{\vec q}\vert \Delta_{n,n} (\vec q) \vert^2 
\exp (-\frac{q^2l^2}{2}) \right. \nonumber \\
& & \left. + 2 \eta (\Delta_{n,n}(0)-\rm{Re} 
(\Delta_{n,n+1}(0)))\right] \nonumber \\
& \equiv & N_\phi N_z g^2 (E_{2d} + \eta E_J).
\label{eq:eleven}
\end{eqnarray}
Here $g = \alpha_H\sqrt{\pi \ell^2 d_0/\beta k_B T}$
and $\eta = (\hbar^2/2m_cd^2)|\alpha_H|^{-1}$ measure the intra-layer and 
inter-layer coupling respectively,
$E_{2d} = N_z^{-1}\sum_n [\rm{sgn} (\alpha_H) \Delta_{n,n}(0)
  + \sum_{\vec q}(\vert \Delta_{n,n} (\vec q) \vert^2/4) 
     \exp (-q^2 \ell^2/2)]$ is the intra-layer energy and 
$E_J = 2N_z^{-1}\eta \sum_n (\Delta_{n,n}(0)-\rm{Re} (\Delta_{n,n+1}(0)))$
is the inter-layer Josephson coupling energy.  For $\alpha_H (T) < 0$ 
the square of the dimensionless 
intra-layer coupling constant ($g^2$) is the ratio of the mean-field-theory 
superconducting condensation energy per-vortex per-layer to the thermal
energy $k_B T$.  For typical high temperature superconductors this ratio 
remains small over a wide range of temperature below the mean field
transition temperature. The dimensionless Josephson  
coupling constant $\eta$ is the ratio of the square of the Gaussian
approximation to the correlation length along the field direction,

\begin{equation}
\xi_H^2 \equiv \frac{ \hbar^2 }{ 2 m_{c} |\alpha_H(T)| } = 
\xi_c^2 \frac{ |\alpha(T)|}{|\alpha_H(T)|}, 
\label{eq:twelve}
\end{equation}
to the square of the layer separation.
Several researchers\cite{glazman,dimcross} have speculated on
possible experimental consequences of the dimensional crossover
from two to three dimensions, naively expected when $\eta$ exceeds $1$. 
We will comment on these issues later.

\section{Numerical Monte Carlo Simulations} 

We performed Monte Carlo simulations of the LLL limit of the 
Lawrence-Doniach model for finite size systems.
The complex expansion coefficients $C_{n,j}$ of the 
order parameter in Eq.~(\ref{eq:four}) were treated as
classical variables in the 
simulations. The real and imaginary parts of $C_{n,j}$ were 
independently incremented by random numbers from a distribution chosen
so that half of all attempted moves were accepted using the standard 
Metropolis algorithm.  Each Monte Carlo step consisted of update attempts for 
the real and imaginary parts of all $N_{\phi} N_z $ coefficients.
We calculated distribution functions and averages for various
quantities of interest, 
including the total Ginzburg-Landau energy, the separate planar and Josephson
coupling contributions to the Ginzburg-Landau energy, interlayer and 
intra-layer local superfluid density correlation functions, and 
correlation functions that are off-diagonal in the layer index.
The simulations were performed for a range of values of the two
dimensionless parameters that characterize the system, $g$ and 
$\eta$, and for a range of system sizes.  We have focused our attention
primarily on simulations at small values of $\eta$, where the 
discreteness of the superconducting layers plays an important role,
because this is the situation of relevance to most high-temperature 
superconductors.   

Most of our Monte Carlo simulations were performed for finite systems 
containing $N_z = 8$ layers and 
$N_{\phi} = 16$, $N_{\phi}=24$, or $N_{\phi}=36$, vortices per layer.
The number of Monte Carlo steps used in a single simulation was 
typically $M = 2 \times 10^6$.  
Our finite size system shapes have been
chosen to accommodate perfect triangular lattices by taking 
\begin{equation}
\frac{L_y}{L_x} = \frac{2}{\sqrt{3}}\times \frac{N_y}{N_x}
\label{eq:thirteen}
\end{equation} 
where $N_y$ and $N_x$ are integers and $N_\phi = N_x \times N_y$.
(We choose $N_x=N_y=4$ for $N_{\phi}=16$; $N_x=6$, $N_y=4$ for 
$N_{\phi}=24$ and $N_x=N_y=6$ for $N_{\phi}=36$).
For all simulations the order parameter was
initialized to the Abrikosov lattice value
and the first $10^4$ Monte Carlo steps were discarded.
We have compared results from different stages of
our Monte Carlo runs to make sure our systems were
well equilibrated.

\begin{figure}
\psfig{file=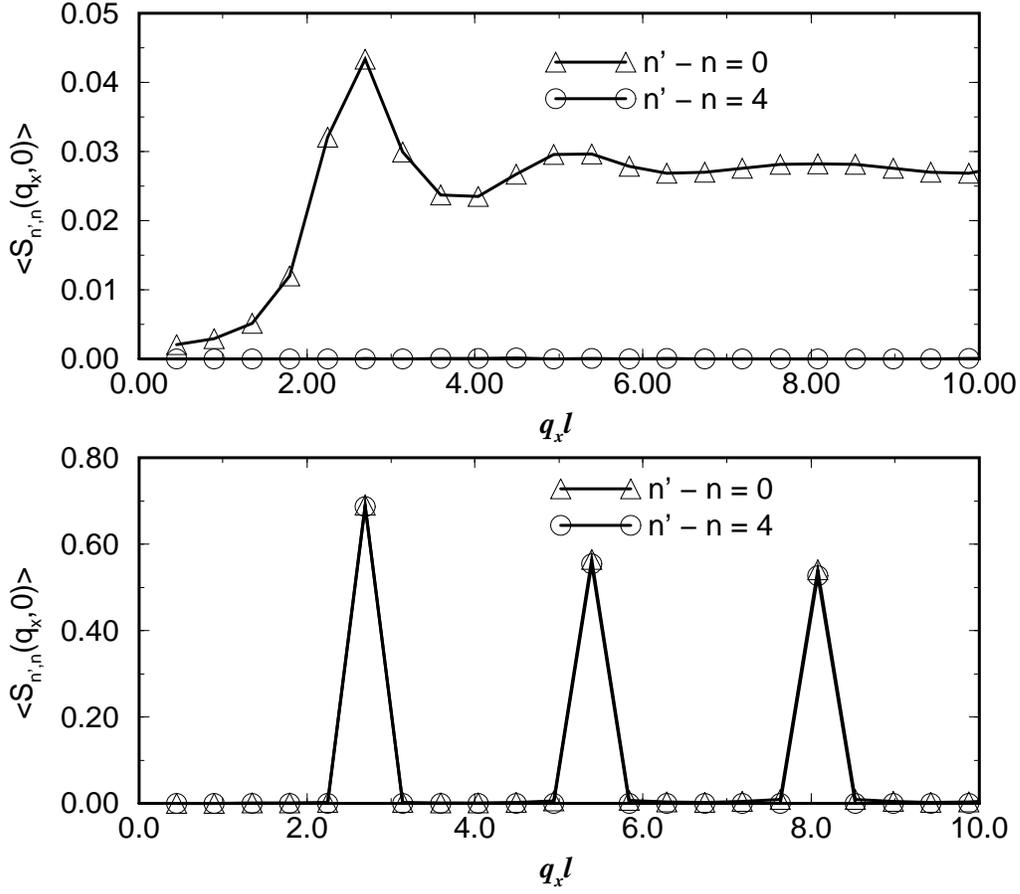,width=15cm}
\caption { 
$\langle S_{n^\prime,n} (q_x, 0) \rangle $
at $\eta = 0.01$ and $g = -\protect\sqrt{10}$ (upper panel),
and $g = -\protect\sqrt{30}$.  These results are 
for a finite system with $N_{\phi} = 36$ and $N_z = 8$
and were obtained by averaging over $10^6 $ Monte-Carlo steps.
}
\label{fig1}
\end{figure}

Some typical results for $ \langle S_{n^\prime,n} (\vec q) \rangle$
are shown in Fig.~\ref{fig1}.  All the results reported here are for 
temperatures below the mean-field transition temperature so that 
$g < 0$ and larger $|g|$ corresponds to lower temperatures.  
In isolated 2D systems, the pancake vortices of the LLL model melt\cite{2d}
at $g \approx -6.6$.  We expect coupling between layers to increase 
the transition temperature.  The results in Fig.~\ref{fig1} show that, for 
this system size, a qualitative change in local superfluid density correlations
occurs between the two temperatures for which results are shown.  At the 
higher 
temperature ($g = - \sqrt{10}$), the peak in $S_{n,n}(q)$ near reciprocal 
inter-vortex distances
demonstrates that the pancake vortices are in a moderately correlated 
liquid state.
There is no evidence of correlations between vortices in different layers,
perhaps not surprisingly in view of the small value for $\eta = 0.01$. 
On the other hand for $g = - \sqrt{30}$, $S_{n',n}(\vec q)$ is nearly 
independent 
of the layer indices, indicating that vortices in different layers 
are highly correlated.  Moreover, $S_{n',n}(\vec q)$, is strongly peaked
at reciprocal lattice vectors, reaching a value that is approximately
$70\%$ of the Abrikosov state value for the nearest neighbor shell.
Apparently, once the vortices are strongly correlated within a 
layer, weak interlayer coupling is sufficient to strongly favor
configurations in which the vortex coordinates are weakly dependent 
on layer index, {\it i.e.} configurations with nearly straight vortex
lines.
 
\begin{figure}
\psfig{file=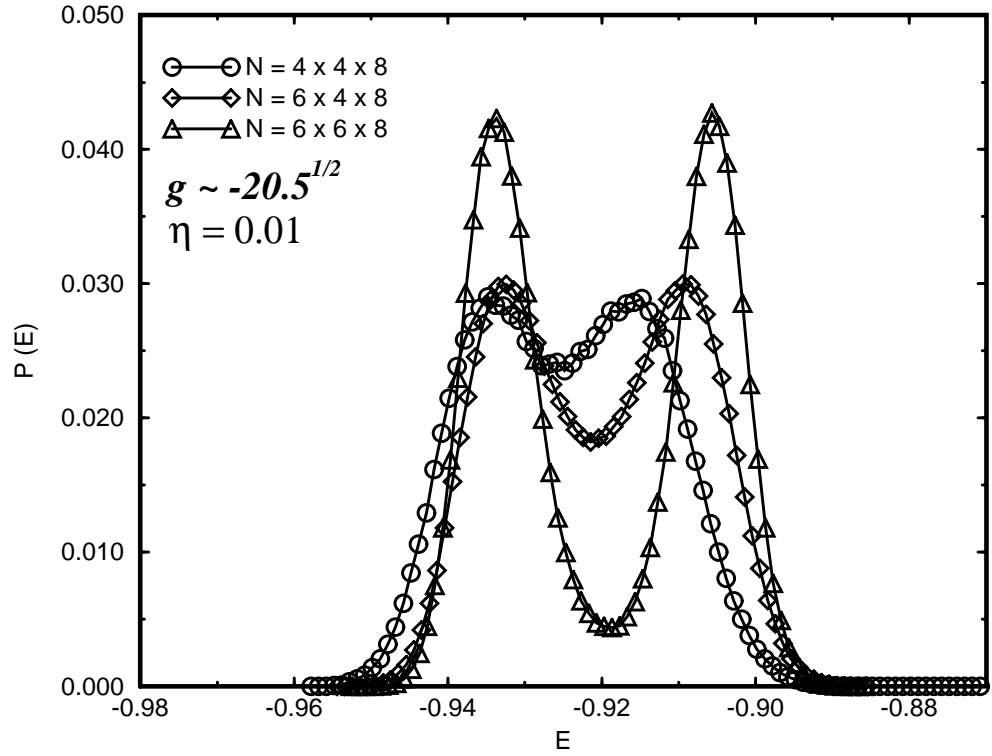,width=15cm,angle=270}
\caption{Landau-Ginzburg energy distribution function at
the finite system phase transition point for various system sizes.
Energies are in units of the mean-field condensation energy,
$N_{\phi} N_z k_B T g^2/ \beta_A$.
}
\label{fig2}
\end{figure}

Our simulations indicate that the transition between these two patterns of 
correlation happens abruptly at what appears to be a first order
phase transition, which for $\eta = 0.01$ occurs for 
$g \approx - \sqrt{20.5}$.
In Fig.~\ref{fig2} we show the probability 
distribution function for the Ginzburg-Landau free energy 
at $\eta = 0.01$ and $g = - \sqrt{20.5}$.  The double-peaked distribution
function with a intermediate minimum which deepens with system size
is strong evidence~\cite{ref:5,ref:6} that a first order phase 
transition occurs 
in the thermodynamic limit in this system.  The logarithm of the ratio of the 
maxima in $P(E)$ to the intervening minimum may be interpreted as 
a free energy barrier between the low temperature vortex lattice state and 
the high temperature vortex liquid state.  The transition will be first order 
if this barrier diverges in the thermodynamic limit.
By this measure, even for this relatively weak Josephson coupling,
the phase transition is much more strongly 
first order than for uncoupled layer systems.  In that case
the free energy barrier grows more slowly with system sizes and 
first appears in the simulations only when the number of vortices
per layer is close to $100$~\cite{2d,kato}.  The energies in 
Fig.~\ref{fig2} are given in units of the mean-field-theory 
condensation energy.  We see that for this value of $\eta$, the Ginzburg-Landau
free-energy at the transition is, in magnitude, about $93.5 \%$ of its 
mean-field value when in the vortex lattice state and about $90.5 \%$ of 
its mean field value when in the liquid state.  It would be tempting 
to identify the separation between the two peak energies in Fig.~\ref{fig2}
with the latent heat.  However, as we discuss later, this identification
would be incorrect because of the temperature dependence of the
parameters of the Ginzburg-Landau free energy.

\begin{figure}
\psfig{file=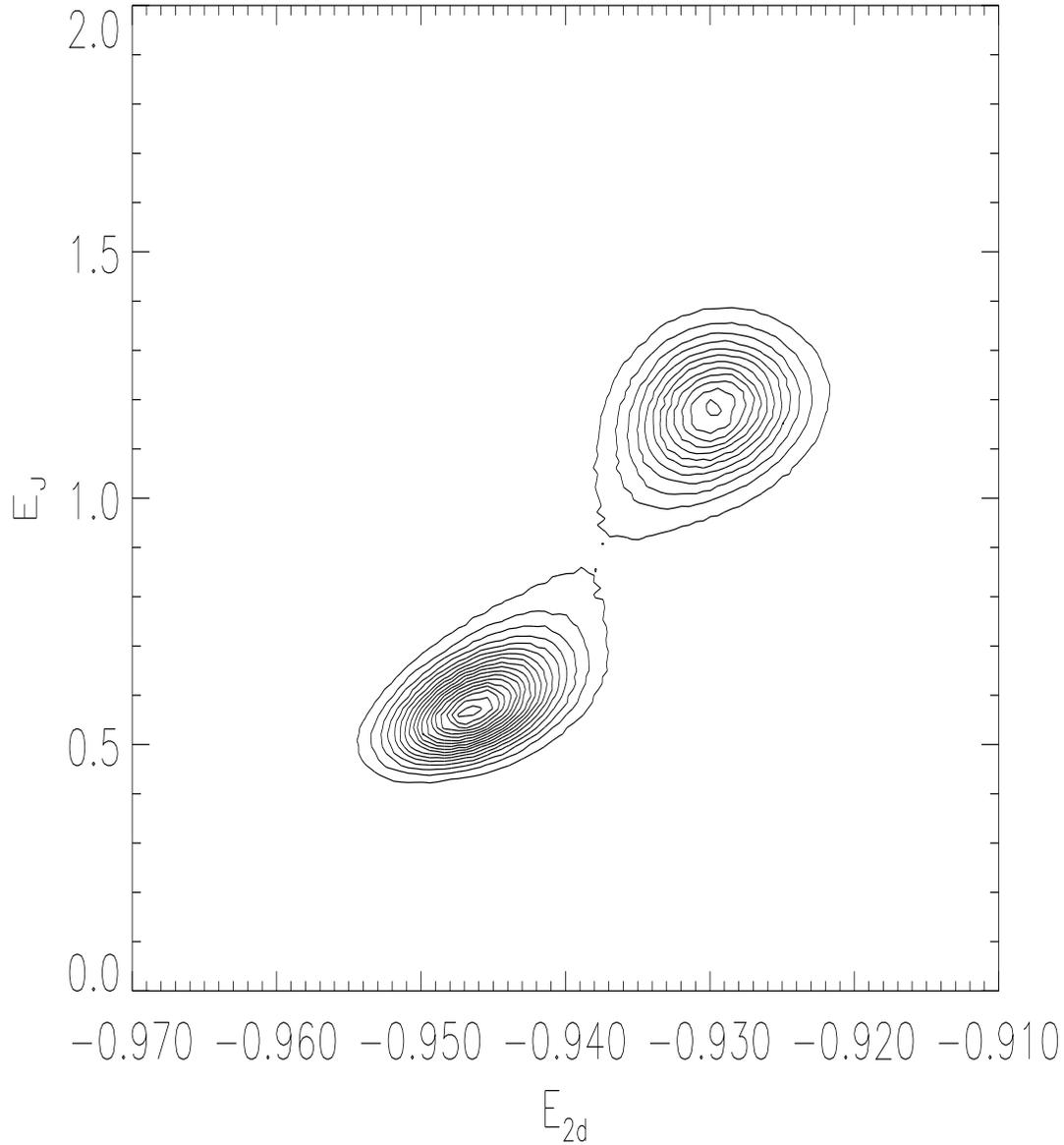,width=14cm,height=16cm}
\vspace*{0.5in}
\caption{
Two dimensional contour plot for the distribution function
of $(E_{2d}, E_J)$ at the finite system phase transition 
point $(\eta_m, g_m) = (20.52, -0.01)$. 
The peak located at smaller $<E_{2d}>$ and $<E_J>$
comes from 3D vortex solid configurations.
The peak located at larger $<E_{2d}>$ and $<E_J>$
comes from vortex liquid configurations.
}
\label{fig3}
\end{figure}

Fig.~\ref{fig3} contains a contour plot of the two-dimensional 
distribution function $P(E_{2d},E_J)$ at the transition.
This picture reinforces the picture of the transition 
as being controlled by a competition between 
planar and Josephson coupling energies,
both of which favor order, and entropic contributions to the free 
energy which favor disorder.  The planar energy decreases because typical 
vortex configurations more closely approximate their 
triangular lattice optimum while Josephson coupling energies 
drop because of increased interlayer correlation.
The distribution function plot
demonstrates that both the planar Ginzburg-Landau energy and the 
Josephson coupling energy are lower in the ordered state.
We have defined $E_J$ so that it is zero if the order parameter
is identical in every layer.  For completely uncorrelated order
parameters in different layers, $\Gamma \equiv E_J /\Delta_0 = 2 $. 
We find that $\Gamma$ has the values $ 0.598$ and $ 1.208 $ in
ordered and disordered 
states respectively at the $\eta = 0.01$ phase transition;
$\Gamma$ drops from approximately $60\%$ to $30\%$ of its 
uncorrelated layer value on going form disordered to ordered states.
This large change should be compared with the more modest 
relative change in the planar condensation energy.
We show later that the latent heat and magnetization discontinuity
at the first order phase transition are related to the changes
in planar condensation energy and Josephson coupling energy that can 
be read off these figures.

\section{Discussion and Comparison With Experiment} 

\subsection{Phase diagram}

We have completed Monte-Carlo simulations of our model 
at a series of values for the two dimensionless parameters 
($ g$ and $\eta$) which characterize it.  
These two parameters can be related to 
superconductor material properties that are measurable,
at least in principle:
\begin{equation}
g^2 = \frac{\Lambda_T(T_{c0})}{\Lambda_{eff}}\frac{T_{c0}}{T}
      (\frac{T_{c0}}{T}-1)\frac{H^{MF}_{c2}(T)}{H}
		(1-\frac{H}{H^{MF}_{c2}(T)})^2
\label{fourteen}
\end{equation}
and
\begin{equation}
-g \eta = \frac{1}{\kappa\gamma^2}(\frac{\Lambda_T\ell^2}{2d^3})^2 = 
   \frac{1.81\times 10^6}{\kappa\gamma^2}(\frac{1}{HTd^4/d_0})^{1/2}
\label{fifteen}
\end{equation}
where $\Lambda_T (T) = \Phi_0^2/16\pi^2k_BT = 2\times 10^8 \AA / T$ is the
thermal fluctuation length, 
$\Lambda_{eff} = 2\lambda_{ab}^2 (0) / d_0 $ is the two-dimensional
(2D) effective penetration 
depth, $\kappa=\lambda/\xi$, and 
$\gamma^2 = m_c/m_{ab}$.  The numerical form of the equation for 
$g \eta$ applies for $H$ in units of Tesla, $T$ in units of Kelvin, 
and $d_0$ and $d$ in units of $\AA$.
Note that $g \eta$ does not vary strongly 
on crossing through the fluctuation regime with varying temperature
or field, and that this product can be quite small for strongly anisotropic
extreme type-II superconductors.  
Where our model is valid, the phase diagrams and appropriately scaled 
physical properties (see below for examples) of all planar superconductors
should become identical when field and temperature are expressed in terms of 
$g$ and $\eta$.  

\begin{figure}
\psfig{file=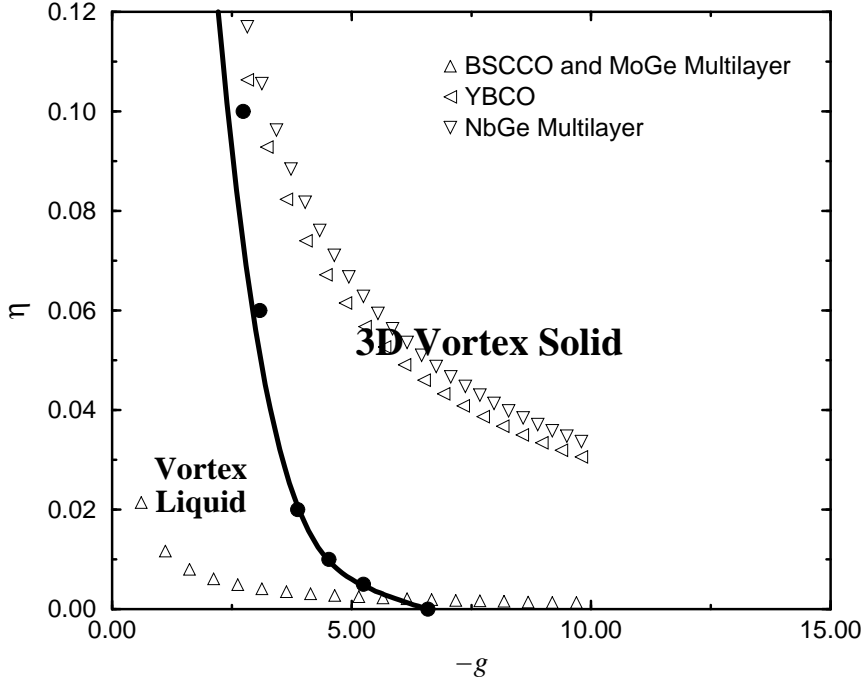,width=15cm,angle=270}
\caption{
The phase diagram $\eta_m = f(g_m)$ in terms of the
two dimensionless parameters $g$ and $\eta$
we use to characterize intra-layer and inter-layer couplings.
Phase transition points located in our simulations 
are indicated by solid circles. 
This phase boundary was obtained by spline fitting using the four 
smallest $\eta$ phase transition points.
The two points at larger $\eta$ are ones at which the system
appears to undergo a phase transition as judged by the temperature
dependence of the correlation functions,
but we are not able to locate the double peak structures
due to limited computation time and small finite system size.  
}
\label{fig4}
\end{figure}

Fig.~\ref{fig4} shows the universal $(\eta, g)$ 
phase diagram we have constructed from Monte Carlo simulation  
data like that discussed in the previous section.
The construction of the phase boundary line, $\eta_m = f(g_m) $, is aided by 
the following Clapeyron\cite{huang} identity:
\begin{equation}
\frac{d\eta_m}{dg_m} = -\frac{2}{g}
[\frac{\delta <E_{2d}>}{\delta <E_J>}+\eta_m].
\label{eq:13}
\end{equation}
Here $\delta < E_{2d}>$ and $\delta < E_J>$ are the respective 
discontinuities across the phase boundary of the 
planar condensation energy and the Josephson coupling energy discussed
in the previous section.  We have used points on the boundary 
and these derivatives to 
obtain the spline fit to the phase boundary line that is indicated in 
Fig.~\ref{fig4}.
The data points used for this fit
were $ (\eta_m, g_m) = (0.0, -6.6)$, $(0.005, -5.2)$,
$(0.01, -4.5)$, $(0.02, -3.9)$ and $d\eta_m/dg_m = 0.024 $ at
$ (\eta_m, g_m) = (0.02, -3.9)$.  
The dimensionless parameters of a particular layered superconductor will move 
through this phase diagram from upper left to lower right on crossing
through the fluctuation regime by decreasing temperature or field.  
The three additional curves in Fig.~\ref{fig4} are lines of constant $g \eta$ 
appropriate for experiments performed at typical fields and temperatures
on various materials.
For the quasi two-dimensional BSCCO family of superconductors the typical
value of $\eta$ at the phase boundary is quite small. 
(At $H = 1 $ Tesla, $T=50$ Kelvin, $\eta= 0.0014$ for BSCCO.) 
On the other hand, for bulk superconductors $g \eta$ at the phase
boundary can easily be larger than $1.8$,
the value of the product at the upper right of the region illustrated
in Fig.~\ref{fig4}. 

\begin{figure}
\psfig{file=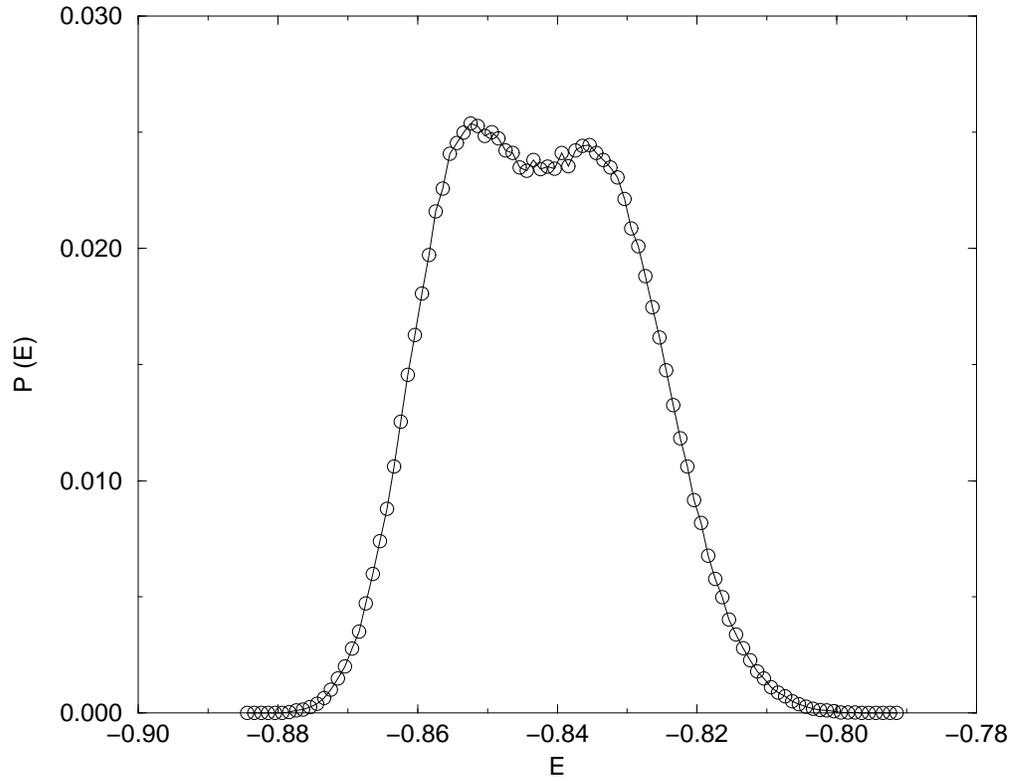,width=15cm,angle=270}
\caption{Landau-Ginzburg energy distribution function near
the finite system phase transition point for $g = -9.5$ and
$\eta = 0.06$.  These results are for $N_z=32$ and 
$N_{\phi} = 16$.  Energies are in units of the mean-field condensation energy,
$N_{\phi} N_z k_B T g^2/ \beta_A$.
}
\label{fig5}
\end{figure}

The two solid dots in Fig.~\ref{fig4} at $\eta = 0.06$ and 
$ \eta = 0.10 $ were not included in determining the spline fit
but lie quite close to the line extrapolated from weaker couplings.  
At these values of $\eta$, our $N_z=8$ simulations found that  
the measured correlation functions changed abruptly with $g$,
but did not uncover the double-peak distribution
in the Ginzburg-Landau free energy that is characteristic
of a first order phase transition.  
At $\eta = 0.06$ more layers are required before the   
simulation begins 
to show a double-peak distribution; in Fig.~\ref{fig5}
we show results obtained at the phase boundary for 
$N_z = 32$ and $N_{\phi}=16$.

As $\eta$ increases, the correlation length along 
the field direction increases, and  
finite $N_z$ effects should increase in 
importance.  At larger values of $\eta$, 
we believe that the phase boundary line would move
toward the right if we increased the number of layers.  
One way to judge the importance of finite size effects is to recognize
that for $\eta \to \infty$
the order parameter will become layer independent for any
finite $N_z$ and our model will reduce to a 
2D model with $ g^2 \to g^2 N_z$.  Since 
for an isolated 2D layer we have found\cite{2d,kato} that 
$g_m (\eta =0) = -6.6$, we expect to find $g_m = -6.6/\sqrt{N_z}$ whenever
the correlation length along the field direction in the fluid phase has 
reached the system size $N_z d$.
For example, for simulations with $N_z = 8$, like
those on which Fig.~\ref{fig4} is based, $g_m$ will never be
less than $ \approx 2.3$.  The phase boundary we find at $\eta =0.12$ is
perilously close to this limit and it is likely that for $N_z =8$ we 
are quite close to the 2D limit at this interlayer 
coupling.~\cite{doublepeakremark}
Even for YBCO it seems likely that more than $N_z = 8$ layers are 
required for a simulation of the phase transition\cite{sasik}.

As mentioned previously, for materials with reasonably narrow fluctuation
regimes,  $ - g \eta $ is essentially constant on sweeping through
the transition regime with either field or temperature.  
Values of $ - g \eta $ for a number of planar superconductors 
at typical fields and temperatures
are indicated in Table~\ref{table1}.
With this in mind we remark that in high temperature superconductors, and 
in other highly anisotropic planar systems, the transition to the vortex 
lattice state
will always occur in the regime where $\eta < 1$ and the discrete nature 
of the superconducting layers is of importance.  To see this we observe that
in the limit of large $\eta$ our model reduces to the LLL model
for continuum three dimensional systems.  In that limit, properties of 
the model are, as in the isolated layer limit, controlled by
a single dimensionless parameter\cite{thouless} 
\begin{equation}
g_{3D}^2 = g^2 \sqrt{\eta} 
\label{eq:sixteen}
\end{equation}
$g_{3D}^2$ is the ratio of the mean-field-theory condensation 
energy, per vortex and per correlation length, to the thermal
energy $k_B T$.   It follows from Eq.~(\ref{eq:sixteen}) that 
for large $\eta$, $\eta_m \propto g_m^{-4}$ in Fig.~\ref{fig4}. 
The phase transition to the Abrikosov state must still be strongly
influenced by fluctuations for $\eta \gg 1$ 
since dimension $D=3$ is still below the naive lower critical
dimension for this problem. 
There is, as yet, no quantitative estimate of the 
value of $g_{3D}$ at which the putative phase transition occurs 
in this 3D limit, but it seems clear that its value 
must at least be larger than unity. 
If this is so, the value of $\eta$ at the phase boundary
must satisfy the inequality $\eta_m < (g \eta)^{4/3}$.  Only 
materials with $g \eta > 1$ can be in the 3D continuum 
limit at the phase boundary.
Our attention here is strictly focused on materials with $ g \eta < 1$.

\subsection{Thermodynamics}

\begin{figure}
\psfig{file=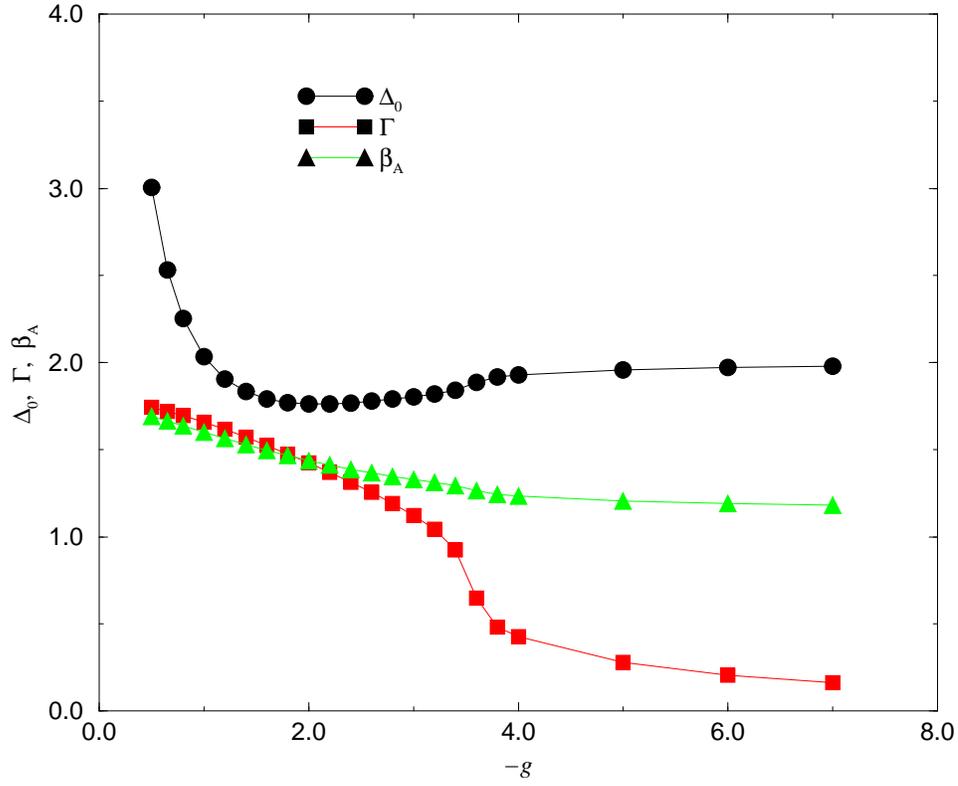,width=15cm,angle=270}
\caption{The $g$ dependence of thermodynamic variables $\Delta_0$,
$\beta_A$ and $\Gamma$ for $g\eta = -0.01$. The calculations are done
for the system size $N_\phi = 16$, $N_z = 8$.}
\label{fig6}
\end{figure}

It is useful to discuss the thermodynamics of the LLL Lawrence-Doniach
model using a microcanonical ensemble approach that we introduced
earlier for isolated layers\cite{abent,russianabent} and we briefly
generalize below.  This approach is
based on the observation that the Ginzburg Landau free energy can be
expressed in terms of a small number of physically meaningful, intensive
thermodynamic variables.  For the present case we introduce three variables: 
$\Delta_0$, the Abrikosov ratio $\beta_A$ 
\begin{equation}
\beta_A \equiv N_z N_{\phi} (2 \pi \ell^2 ) \frac{ \sum_n \int d^2 \vec r |
 \Psi_n (\vec r)|^4}{ (\sum_n \int d^2 \vec r |\Psi_n (\vec r)|^2)^2}
\label{eq:new1}
\end{equation} 
and a Josephson coupling parameter 
\begin{equation}
\Gamma \equiv \frac{\sum_n\int d^2\vec r|\Psi_{n+1}(\vec r)-\Psi_n(\vec r)|^2}
		{\sum_n \int d^2 \vec r |\Psi_n (\vec r) |^2 }.
\label{eq:new2}
\end{equation}
The Abrikosov ratio is small for smooth order parameters whose 
zeroes are relatively evenly spaced in each plane; 
it's minimum value $ \beta_A = \beta_{\bigtriangleup}$ is achieved when the 
vortices are placed on a triangular lattice~\cite{abref} and
it reaches the maximum value $\beta_A = 2 $ for the vortex gas state.  
$\Gamma$ has minimum and maximum values $0$ and $2$ 
for the Abrikosov lattice and vortex gas states respectively. 
Fluctuations in these intensive thermodynamic variables will vanish
in the thermodynamic limit.
In Fig.~\ref{fig6}, we plot $\Delta_0$, $\beta_A$, and $\Gamma$
as a function $g$ for a finite system with 
fixed $-g\eta = 0.1$.   The melting transition for this
value of $g \eta$ occurs near $g=-3.6$ where 
the Josephson coupling parameter decreases relatively abruptly.  

With these definitions the LLL Lawrence-Doniach Ginzburg-Landau free energy 
can be expressed in the form:
\begin{equation}
F_{GL} = k_B T N_{\phi} N_{z} f 
\label{eq:new3}
\end{equation}
where 
\begin{equation}
f = g^2 (sgn(g)\Delta_0 +\frac{\beta_A \Delta_0^2}{4}+\eta \Delta_0 \Gamma).
\label{new4}
\end{equation} 
The free energy of the LLL Lawrence-Doniach model can be expressed as the 
sum of $F_{GL}$ and an entropic contribution proportional to the log of 
the volume in the model's phase space with given values of 
the three variables.  Noting that both $\beta_A$ and $\Gamma$ have been 
defined so that they are invariant under scale changes of the order parameter,
it follows\cite{abent} that the free energy of the model is 
\begin{equation} 
F = k_B T N_{\phi} N_z [ f  - \ln (\Delta_0) -s(\beta_A,\Gamma)].
\label{eq:new5}
\end{equation} 
where $s(\beta_A,\Gamma)$ is an entropy function that completely 
specifies the thermodynamics of the model.
The equilibrium values of the thermodynamic variables $\Delta_0$, $\beta_A$,
and $\Gamma$ and the free energy are determined at specified values of the 
dimensionless model parameters, $g$ and $\eta$, by minimizing the right hand 
side of Eq.~(\ref{eq:new5}).  For example, minimizing with respect to 
$\Delta_0$, we find that for $g < 0$, {\it i.e.}, below the mean-field 
transition temperature, 
\begin{equation}
\beta_A = \frac{2(1-\eta \Gamma)}{\Delta_0}+\frac{2}{\Delta_0^2 g^2}.
\label{new7}
\end{equation} 
This constraint on the three thermodynamic variables we 
have introduced will be used below to 
simplify expressions for the latent heat and the
magnetization discontinuity at the model's first 
order phase transition.  Minimizing with respect to $\Gamma$ or  
$\beta_A$ yields a relationship that involves the entropy function.
It is easy to show that $s(\beta,\Gamma)$ reaches its maximum value 
at the high temperature equilibrium values $\beta_A =2$ and $\Gamma =2$;
however the properties of this function
have not yet been thoroughly investigated.  
Minimizing with the entropic terms in 
Eq.~(\ref{eq:new5}) ignored produces mean field theory. 

The thermodynamic signature of a first order phase transition between
the disordered vortex liquid phase and the ordered vortex solid
phase is the presence of a latent 
heat and a discontinuity in the magnetization of the system.  
The latent heat is proportional to the decrease of entropy on 
entering the ordered state.  In our LLL Lawrence-Doniach model 
the entropy can be expressed in the following form
\begin{equation}
S = \frac{\langle F_{GL} \rangle-F}{T} - \langle \frac{\partial
 F_{GL}}{\partial T} \rangle .
\label{new8}
\end{equation}
Since the free energy is continuous at the transition, the 
latent heat has contributions from the discontinuity in 
$\langle F_{GL} \rangle $ and the discontinuity in 
the last term on the right hand side of Eq.~(\ref{new8}).
The first contribution to the latent heat is what would have been 
read off the separation between the two-peaks in the 
distribution function of $F_{GL}$ near the transition
and represents a contribution from the order parameter phase 
space volume.  The second term is due to the temperature dependence of 
$F_{GL}$ and represents a contribution from fluctuations 
on a microscopic level for a given order parameter function.
This contribution to the entropy is proportional to the 
magnitude of the average local superfluid density if we 
assume that only $\alpha$ depends on $T$.
We remark that the possibility of a contribution to the 
entropy of this type
is discarded from the beginning in a frustrated $XY$ or London
model description of 
the phase transition\cite{hetzel,chenteitel,sudbo}, since the magnitude 
of the order parameter is implicitly held constant
and no attempt is normally made to estimate the temperature
dependence of the parameters of this effective model.

Using Eqs.~(\ref{new4}), ~(\ref{new7}) and ~(\ref{new8}), the latent heat 
can be expressed in terms of material parameters  
and discontinuities in intensive thermodynamic variables across the transition.
\begin{equation}
T_m\delta S = N_\phi N_z k_B T_m g_m^2 [-\frac{\delta <\Delta_0>}{2} + 
                             \frac{\eta}{2} \delta <E_J>
               + \frac{T_m}{H_{c2}^{MF}(T_m)-H_m} 
                 \frac{dH_{c2}^{MF}}{dT}\delta <\Delta_0>].
\label{eq:14}
\end{equation}
(Recall that $E_J = \Delta_0 \Gamma$.)  The magnetization in the LLL model  
comes entirely from the dependence of the model parameters on field and is 
proportional to $\Delta_0$:   
\begin{equation}
\frac{4 \pi M }{H} = -\frac{k_BTg^2}{2\pi\ell^2d_0}\times
                      \frac{\Delta_0}{H_{c2}^{MF}-H}
					= -\frac{(H_{c2}^{MF}-H)\Delta_0}{4\kappa^2H}.
\label{eq:15}
\end{equation}
Since $\Delta_0$ increases on freezing, the internal magnetic field
decreases and the vortex lattice therefore expands slightly.
This expansion will always occur in the LLL limit.\cite{likeice} 
Eq.~(\ref{eq:15}) is valid for extreme type-II superconductors 
in the LLL regime since  
the screening current gives rise to a small correction to
the internal field which is approximately uniform spatially.
There is no need
to distinguish between internal and external fields on the right hand
side of this equation.  (This is the same property which has allowed us 
to ignore thermal fluctuations in the internal magnetic field.)   

\begin{figure}
\psfig{file=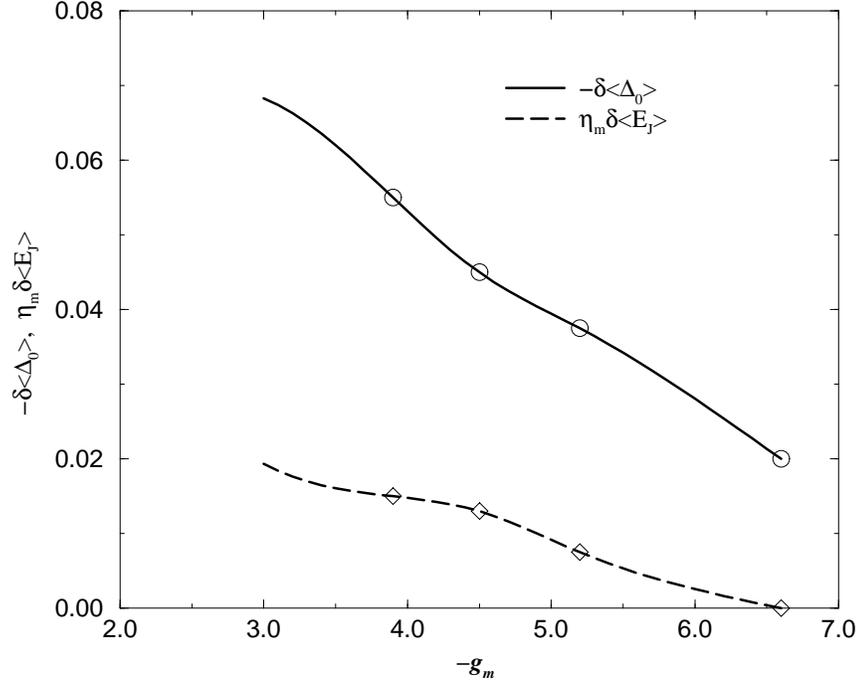,width=15cm,angle=270}
\caption{
The $g_m$ dependence of the discontinuity for $-\Delta_0$ and $E_J$ along
the phase boundary between vortex liquid state and vortex solid state.
}
\label{fig7}
\end{figure}

Our calculations in principle allow theory and experiment to be quantitatively
compared for any strongly anisotropic layered superconductor.
For a specific superconducting material, with known $H_{c2}^{MF}(T)$,
$\kappa$, $\gamma$, $d$, and $d_0$, 
the point $(g_m,\eta_m)$ in parameter space where an apparent 
first order phase transition occurs can be calculated from 
field strength and temperature using Eq.~(\ref{fourteen}) and 
Eq.~(\ref{fifteen}).
The applicability of the LLL model can be judged from the proximity 
of the melting point to the estimates given here in Fig.~\ref{fig4}.
Our estimate for the latent heat and the magnetization discontinuity can
then be calculated using Eq.~(\ref{eq:14}) and Eq.~(\ref{eq:15}) and 
the discontinuities in $E_J$ and $\Delta_0$ plotted in 
Fig.~\ref{fig7}.  
We remark that where the LLL limit is valid,
the relative change in the internal magnetic field for strong type-II
superconductors at the transition
is always small, less than $\sim 1/ (200\kappa^2)$.
The entropy jump per vortex per layer can be substantially larger
and may be estimated to be:
\begin{equation}
\frac{\delta S}{k_B N_{\phi} N_z} \sim \frac{g_m^2 
\delta \langle \Delta_0 \rangle} {(1-T_{c0}/T_m)(1-H/H_{c2}^{MF}(T_m))} 
\label{new10}
\end{equation}
The entropy jump per vortex, per layer can be much larger than the value 
estimated in $XY$ model simulations 
~\cite{hetzel} in which the dominant
contribution to Eq.~(\ref{eq:14}) is absent.  We compare 
our results with some recent experiments in a subsequent subsection.  

\subsection{Off-diagonal long-range-order}

\begin{figure}
\psfig{file=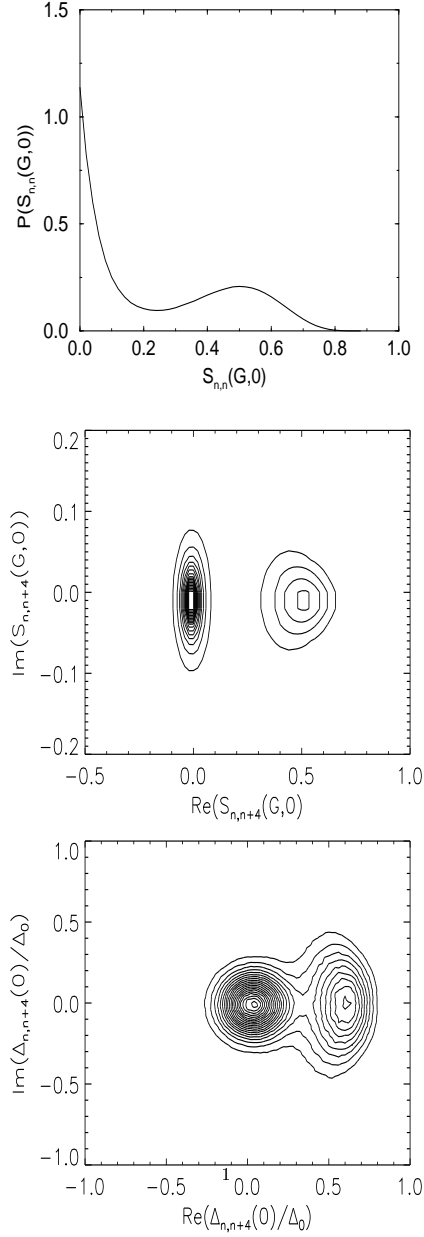,height=19.0cm}

\caption{The top panel shows the distribution function for 
$S_{n,n} (G,0)$;
the middle panel is a contour plot of the distribution function for 
        $S_{n,n+4} (G,0)$ in the complex plane; the bottom panel 
        is a contour plot of the distribution function for 
	  $ \Delta_{n,n+4} (\vec q = 0)/\Delta_0$ in the complex plane.
        All plots are at the melting transition point 
	  of a finite system with $N = 6\times 6\times 8$, 
   $g = -\protect\sqrt{20.5}$, and $\eta = 0.01$. $\vec G$ is a member
of the first shell of reciprocal lattice vectors of the
Abrikosov lattice. 
}
\label{fig8}
\end{figure}

We now turn our attention to the evidence for superconductivity 
and off-diagonal order that comes from simulations.   
In Fig.~\ref{fig8} we compare results for local
superfluid density correlation functions and
off-diagonal correlation functions at the phase transition 
point where contributions due to both ordered and disordered
states contribute.  The peak in the $\Delta_{n,n+4}({\vec q} =0)$ 
distribution function\cite{remark1} centered on a non-zero 
value comes from the same ordered state configurations that
give peaks in the probability distribution functions for $S_{n,n}(G,0)$ 
and $S_{n,n+4}(G,0)$.  Note that phase fluctuations along 
the field direction cause this quantity to have a 
non-zero imaginary part, and that the peak in the 
distribution function has a greater width in the imaginary 
direction than in the real direction, indicating that phase
fluctuations have a greater importance than amplitude 
fluctuations.  There is no evidence of these off diagonal
correlations above the vortex lattice melting transition.
This result contrasts with $XY$ model studies which find 
a superconducting vortex line liquid state that has phase 
coherence along the field direction without  
broken translational symmetry~\cite{chenteitel,sudbo}. 
The explanation for this difference might originate 
in the fact that no clearly defined region which is 
outside of the vortex cores exists in the LLL approximation,
making liquid state vortex position fluctuations more 
effective in destroying coherence along the field.
The absence of such a phase in the LLL
approximation does not rule out the possibility that it
exists at weaker magnetic fields.\cite{tesanovic2}

The concept of ODLRO and has long played a fundamental role in 
the theory of superfluidity and superconductivity.\cite{yang}
Only recently\cite{glazman,odlro} however, have attempts been made to apply
this concept to superconductors in a magnetic field.
These efforts have led to some confusion and controversy, 
partly due to the gauge dependence of Cooper pair 
correlation functions in a magnetic field. 
The relationship is an interesting one in GL models generally and in 
LLL models in particular.~\cite{tesanovic1} 
(The relationship is partially obscured in $XY$ and 
London models for vortex states.)  
In the absence of a field, superconductivity in 3D systems 
is associated with Bose condensation of Cooper pairs, 
{\i.e.} with macroscopic occupation of their zero momentum
state.  This properties immediately implies ODLRO
, {\it i.e.} that  
\begin{equation}
\lim_{\vec R \to \infty} \langle \bar
 \Psi (\vec r + \vec R) \Psi (\vec r) \rangle \ne 0.
\label{eq:odlro1}
\end{equation}
In a magnetic field, Cooper pair states are 
labeled by a momentum along the field direction (restricted to the 
appropriate Brillouin zone for the Lawrence-Doniach model), 
by a Landau level index, and by a label that counts the $N_{\phi}$ states
within a Landau level.  Superconductivity is presumably
associated with the macroscopic occupation of a Cooper pair state
with $k_z=0$ and with the lowest Landau level index, but it 
less obvious what state should be occupied within a Landau level.

In order to avoid the requirement for prior knowledge about 
the state in which Bose condensation has occurred,
we propose defining off-diagonal long-range-order in a 
magnetic field in terms of the eigenvalues, $\rho_{\alpha}$,
of the Cooper pair density matrix.  For the case of the
Lawrence-Doniach model the density matrix 
takes the form
\begin{equation} 
\langle \bar \Psi_{n'} (\vec r^{\, \prime}) \Psi_n (\vec r) \rangle  
= \sum_{\alpha} \rho_{\alpha}  \bar \phi^{\alpha}_{n'} (\vec r^{\, \prime})
\phi^{\alpha}_n (\vec r)
\label{eq:odlro2}
\end{equation} 
where $\phi^{\alpha}$ is a normalized eigenfunction of the density 
matrix. ($\sum_n \int d^2 |\phi^{\alpha}_{n}(\vec r) |^2 =1$.) 
$\sum_{\alpha} \rho_{\alpha}$ will be proportional to the system
size whenever superconducting fluctuations are present in the system.
For example, in our finite-size LLL Lawrence-Doniach simulations,
\begin{equation}
\sum_{\alpha} \rho_{\alpha}
= \frac{ N_{\phi} N_z |\alpha_H|  \pi \ell^2} { \beta} \Delta_0. 
\label{eq:odlro3}
\end{equation}
In a magnetic field, a GL model may be said to possess 
ODLRO when at least one eigenvalue of the density matrix 
is extensive and the corresponding eigenfunction is extended.
We remark that in a magnetic field 
this criterion cannot be satisfied
unless the system breaks translational symmetry\cite{tesanovic1} 
either spontaneously, or as a consequence of pinning inhomogeneities. 
Convincing arguments can be advanced\cite{tobepub} in support 
of the view that superconductivity, both parallel and perpendicular to
the field, will occur if\cite{pinningcaveat}
the system has this ODLRO property.

\begin{figure}
\psfig{file=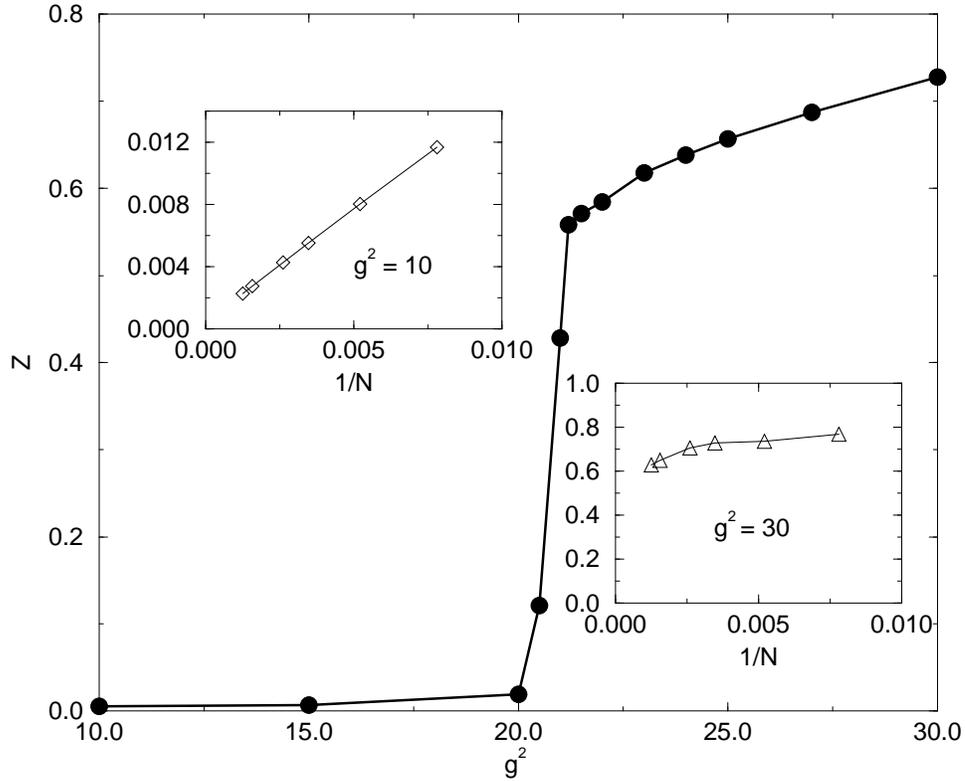,width=15cm,angle=270}
\caption{
The ratio of the largest eigenvalue of the density matrix 
to the sum of all eigenvalues, $Z$,  as a function of $g$ 
at $\eta g = -0.045 $.
For this value of $g \eta$ a first order melting transition 
appears to occur at $g \sim -\sqrt{20.5}$.
The size dependence of $Z$ at particular temperatures on 
the vortex solid and vortex liquid sides of the phase 
transition are shown in the right and left insets respectively.} 
\label{fig9}
\end{figure}

In Abrikosov's mean-field theory, translational symmetry is broken 
to form a triangular vortex lattice, and the density matrix has 
a single extensive non-zero eigenvalue.  When fluctuations are 
included ODLRO will survive if one eigenvalue of the density matrix 
remains macroscopically larger than all others.  This quantity is 
the analog for GL models of the condensate fraction in a boson particle
superfluid.  In Fig.~\ref{fig9}
we plot $Z$, defined as the ratio of the largest
eigenvalue of the density matrix 
to the sum of all eigenvalues, as a function of $g$ at $\eta g = -0.045 $.
For this value of $\eta g$, a first order melting transition
occurs at $g \sim -\sqrt{20.52}$.
The insets in this figure show the systems size dependence of 
the `condensate fraction' for temperatures on the 
vortex solid and vortex liquid sides of the phase transition.
From Fig.~\ref{fig9}  we see that our calculations indicate that
ODLRO occurs simultaneously with broken translational symmetry, at
least when the LLL approximation is valid.

\subsection{Comparing with experiment}

Since the discovery of high temperature superconductivity, 
the superconducting transition in a magnetic field has received 
considerable experimental attention.  It is
abundantly clear\cite{nikulov} that the mean field phase transition,
particularly as seen in specific heat and magnetization measurements,
is spread over a wide fluctuation regime and looks broadly
similar to what would be expected if the system were below its 
lower critical dimension and no true phase transition occurred.
However, according to $XY$ model
simulations\cite{chenteitel,sudbo}
and this and other simulations of the $LLL$ model, there {\it is}
a weak first order phase transition which occurs well below the
mean-field transition temperature and is   
accompanied by a relatively
small latent heat and magnetization discontinuity. 
In this section we compare some of the more recent 
experimental studies\cite{exp1,exp2,exp3,exp4,exp5}
of systems that are free of pinning inhomogeneities 
with our simulations.  In making this comparison we are 
seeking quantitative agreement which would indicate that
the LLL model provides a good description of the experiment
in question. 

We first compare with experimental studies\cite{exp1,exp2,exp4,exp5}
of YBCO which have identified a first
order phase transition in the field and temperature range
$(H_m, T_m) = $ (0--6 Tesla, 93--81 K)
revealed~\cite{exp4,exp5} by a 
small reversible discontinuous increase in the diamagnetism 
of the system with decreasing temperature.  
The latent heat of the transition can then be determined 
from the temperature dependence of the critical field using 
the appropriate Clapeyron equation.  Between $2$ {\rm Tesla} and 
$ 5 $ {\rm Tesla}, the magnetization jump increases from
$\sim 0.012 $ {\rm Gauss} to $\sim 0.025 $ {\rm Gauss} so that the 
relative decrease in the internal magnetic field,
$ 4 \pi \Delta M / H $, varies slowly and is close to 
$7 \times 10^{-6}$.
Over the same field range, the latent heat per vortex per double-layer of 
YBCO is fairly constant and close to $0.65 k_BT_m$.  
From Table 1 we see that in YBCO $g \eta$ is $\sim -0.3$ at 
$ H = 5 $ {\rm Tesla} and varies as $ H^{-1}$ 
in the temperature interval of interest.
Thus the highest fields studied in the
experiment are slightly outside the portion of phase diagram we
explored here.  Nevertheless, the experimental melting phase diagram
for fields larger than $2$ {\rm Tesla} is consistent with the 
extrapolation of our phase diagram to larger $\eta$.  
This claim is reinforced by the LLL Lawrence-Doniach simulations of 
\v{S}\'{a}\v{s}ik and Stroud which specifically
modeled YBCO and found excellent agreement with the measured 
phase boundary.~\cite{sasik2}
Using material parameters 
from Table 1 for YBCO, $dH_{c2}^{MF}/dT=-2.0$ {\rm Tesla/K} 
and the jump in $\Delta_0$ in Fig.~\ref{fig7}
gives $ 4 \pi \Delta M / H \sim 8.9 \times 10^{-6} $ at $H = 5 $ {\rm Tesla}.
Similarly the latent heat calculated from 
Eq.~(\ref{eq:14}) is $0.58 k_B T_m$ per-vortex per double-layer;
both values agree well
with experiment.  For YBCO our results confirm the 
excellent agreement with experiment found 
by \v{S}\'{a}\v{s}ik and Stroud\cite{sasik2} for 
the magnetization discontinuity.
However, these authors
seriously underestimated the latent heat by neglecting 
the last term in Eq.~(\ref{eq:14}) which contributes $\sim 90\%$  
of the total value.  We remark that the latent heat comes 
primarily from the change in entropy content at microscopic
length scales associated with a change in the magnitude of
the superconducting order parameter and not from changes in
the entropy content of vortex configurations.  
In the light of this remarkable quantitative agreement between
theory and experiment, it seems clear that above $H = 2$ {\rm Tesla} 
the phase transition cannot be reliably described by London 
or XY models.

We now turn to the experimental studies\cite{exp3} of the much more
anisotropic BSCCO system where first order phase transitions have 
been found in a much weaker field range: 
$(H_m, T_m) = $ (0--380 Gauss, 90--40 K).
The latent heat at the transition is again $\sim 1.0 k_BT_m $ per vortex
per layer and the jump in the internal magnetic field ($ 4 \pi \Delta
M$) on freezing is $\sim 0.3 G$.
Converting $(H_m, T_m)$ to our dimensionless coupling parameters $g_m$
and $\eta_m$,  we find $g_m$ values up to $\sim 100$ 
for BSCCO in the ranges of fields probed experimentally. 
This is clearly inconsistent with our LLL numerical results 
for the melting temperature.  We have concluded that in
BSCCO, weak interlayer coupling contributes to the high transition
temperatures and relatively low superfluid densities present 
in all high $T_c$ superconductors, resulting in  
fluctuations that are so strong that the 
melting transition occurs outside the validity range 
of the LLL model, even at temperatures quite far from the 
zero-field transition temperature.  
In accepting this conclusion, it is important to realize that 
the largest part of the entropy reduction associated 
with ordering appears in the specific heat and in the 
magnetization at temperatures centered on the mean-field
transition temperature.  At moderate fields,
it has already been convincingly\cite{twodlll} demonstrated that 
thermodynamic properties across these broad fluctuation
regimes are well described by the LLL model; the melting
transition occurs near the low temperature extreme of 
the fluctuation regime and it is only here that the LLL
approximation fails for BSCCO and similar strongly
anisotropic systems.

\section{Summary}

We have presented a discussion of the physics of vortex states
of layered superconductors based on Monte Carlo simulations 
of the lowest Landau level limit of the Lawrence-Doniach
model.  The Lawrence-Doniach model provides a realistic 
description of these systems and may be used to confront 
theory and experiment on a quantitative level.  The lowest
Landau level limit of the model is valid at temperatures
close to the mean field transition and at stronger fields.
In this limit we find that the model has a single first 
order phase transition between a high temperature phase with
no broken symmetries and a low-temperature phase with 
broken translational symmetry and off-diagonal long-range-order,
which we discuss in terms of the eigenvalue distribution of 
the Cooper pair density matrix. 

Although it has been difficult to accurately delimit the validity
of the lowest Landau level constraint on purely theoretical grounds,
quantitative comparison with experiment can provide a 
pragmatic criterion.  With this motivation, we have 
determined a universal phase diagram for this limit of 
the Lawrence-Doniach model in terms of two dimensionless
parameters which can be converted into field and temperature
using the material parameters of a specific layered 
superconductor.  Our numerical simulations have focused 
on the strongly anisotropic limit where the discreteness 
of the superconducting layers plays a role.  
Comparison with experiments indicates that the LLL 
approximation is accurate along the phase boundary for fields
above $\sim 2 $ {\rm Tesla} in YBCO, but that it is not accurate
at more strongly suppressed transition fields in BSCCO 
systems.  In YBCO we find good agreement between
experiment and simulations for both magnetization jumps and 
latent heats associated with the first order phase 
transition.  We conclude that the lowest Landau
level limit of the Lawrence-Doniach model provides a
quantitatively reliable description of the phase transitions
studied experimentally in these systems.  In this model 
the latent heat at the phase transition is primarily due
to entropy changes on microscopic length scales which are  
captured by the temperature dependence of the model parameters and 
are not accurately modeled in XY or London descriptions.

The authors acknowledge informative conversations with 
G. Blatter, A. Dorsey,
S.M. Girvin, D.A. Huse, A.E. Koshelev, S. Teitel, Z. Te\v{s}anovi\'{c},
R. \v{S}\'{a}\v{s}ik, D. Stroud and A. V. Nikulov
This work was supported by the Midwest
Superconductivity Consortium through D.O.E.
grant no. DE-FG-02-90ER45427.

\newpage

\begin{table}
\caption{Material parameters and 
corresponding $g\eta$ at typical temperatures and magnetic fields
for several superconductor systems with strong planar anisotropy.}

\begin{tabular}{||c|l|l|l|l|l|c||}
Material & $\kappa$ & $\gamma$ & $d_0$ (\AA) & $d$ (\AA) & 
                $\kappa \gamma^2d^2d_0^{-1/2}$ & $g\eta$ \\
\tableline
BSCCO~\cite{exp3} & 100 & 55 & 4 & 15 & $3.40\times 10^7$ &
          0.0075 (at $H=1$ T, $T=50.0$ K) \\
\tableline
YBCO~\cite{exp5} & 65 & 7.7  & 3 & 11.4 & $2.89\times 10^5$ & 
          0.30 (at $H=5 $ T, $T=82.5$ K) \\
\tableline
MoGe~\cite{MoGe} & 140 & 22.4 & 60 & 125 & $1.41\times 10^8$& 
          0.013 (at $H=1$ T, $T=1$ K) \\
\tableline
NbGe~\cite{NbGe} & 83.4 & 4.6 & 180 & 202 & $5.37\times 10^6$& 
          0.33 (at $H=1$ T, $T=1$ K) \\
\end{tabular}
\label{table1}
\end{table}

\newpage 

\begin{references}
\bibitem{validitycaveat} The approximation will often be accurate
over a substantial portion of the fluctuation regime
and describe the overall shape of magnetization and specific heat 
curves even when it is not accurate at the
melting phase transition.
\bibitem{doublepeakremark} If so, we would not expect to see a double
peak since for 2D systems these do not become obvious until the number of 
vortices approaches 100.
\bibitem{remark1} We look at off-diagonal correlations at a separation
of four layers since that is the largest layer separation
in an $N_z = 8$ layer system with periodic boundary conditions.
\bibitem{pinningcaveat} Pinning is required for superconductivity 
perpendicular to the field, but its strength can vanish in the 
thermodynamic limit.
\end{references}
\end{document}